\documentclass[lettersize]{IEEEtran}
\usepackage{amsmath,amsfonts}
\usepackage{algorithmic}
\usepackage{algorithm}
\usepackage{array}
\usepackage[caption=false,font=normalsize,labelfont=sf,textfont=sf]{subfig}
\usepackage{textcomp}
\usepackage{stfloats}
\usepackage{xcolor}
\usepackage{url}
\usepackage{verbatim}
\usepackage{graphicx}
\usepackage{cite}
\usepackage{bytefield}
\usepackage{hyperref}
\usepackage{listings}
\usepackage{listings-rust}
\usepackage{float}
\usepackage{tabularx}
\usepackage{multirow}
\usepackage{msc}
\begin{document}

\let\origtextcolor\textcolor
\let\origcolor\color

\newcommand{\DisableColors}{
  \renewcommand{\textcolor}[2]{##2}
  \renewcommand{\color}[1]{}
}

\newcommand{\EnableColors}{
  \let\textcolor\origtextcolor
  \let\color\origcolor
}

\DisableColors

\title{Extensible Post Quantum Cryptography Based Authentication}

\author{
    Homer A. Riva-Cambrin, Rahul Singh, Sanju Lama, Garnette R. Sutherland
\thanks{All authors are affiliated with Project neuroArm, Dept. Of Clinical Sciences, Cumming School of Medicine, University of Calgary.}}



\maketitle

\begin{abstract}
Cryptography underpins the security of modern digital infrastructure, from cloud services to health data. However, many widely deployed systems will become vulnerable after the advent of scalable quantum computing. Although quantum-safe cryptographic primitives have been developed, such as lattice-based digital signature algorithms (DSAs) and key encapsulation mechanisms (KEMs), their unique structural and performance characteristics make them unsuitable for existing protocols. In this work, we introduce a quantum-safe single-shot protocol for machine-to-machine authentication and authorization that is specifically designed to leverage the strengths of lattice-based DSAs and KEMs. Operating entirely over insecure channels, this protocol enables the forward-secure establishment of tokens in constrained environments. By demonstrating how new quantum-safe cryptographic primitives can be incorporated into secure systems, this study lays the groundwork for scalable, resilient, and future-proof identity infrastructures in a quantum-enabled world.
\end{abstract}


\section{Introduction}
\IEEEPARstart{W}{hen} clients access services that contain restricted information, authorization is essential to limit access to intended users. The security of these measures is of special importance for privacy-focused disciplines such as medicine, where patient or health information must be closely guarded \cite{NADHAN2024105511}. Traditionally, this was achieved by sending client credentials (i.e., a username and password) with each request. However, this required the service to store passwords and inherited the well-known weaknesses of password-based authorization like frequent user error \cite{rfc6749}.

A more modern approach is token-based authorization such as OAuth 2.0 \cite{rfc6749,ietf-oauth-v2-1-11}. In OAuth 2.0, a client authenticates with an authorization server to receive an access token, which is used to authorize subsequent requests. Since OAuth 2.0 is not suited for authentication \cite{rfc6749}, protocols like OpenID Connect (OIDC) \cite{openidconnect} extend OAuth 2.0 to provide authentication. The OIDC protocol uses a specific type of token called a JSON Web Token (JWT) \cite{rfc7519}, which rely on digital signatures \cite{saha2016comprehensive}. In a digital signature scheme, a signer holds a private key and makes their public key available to all. Messages signed by the private key can be verified by the corresponding public key, confirming that the private key holder did indeed sign the message. In addition, these signatures also enable integrity verification: if the message is modified, the signature is no longer valid. However, since these tokens carry the signature with each request, they tend to be large, increasing network bandwidth usage.

The emergence of quantum computing presents a new challenge to current cryptographic protocols \cite{bernstein_post-quantum_2017}.  \textcolor{red}{Although quantum computing has allowed for the development of protocols that leverage quantum properties for authentication \cite{lim_quantum_2023, chen_quantum_2023}, most of today's networks are classical networks that cannot implement these techniques in the near future.  Classical cryptographic methods that secure these networks, such as elliptic curve cryptography \cite{Wohlwend2016ELLIPTICCC} and RSA, are threatened \cite{bernstein_post-quantum_2017, Baseri_2024} as they can be broken with Shor's algorithm \cite{Shor_1997}.} These algorithms underpin much of modern secure communication, such as web traffic and financial systems \textcolor{red}{and thus there is a need for post-quantum authentication, that is, authentication that can protect our current infrastructure from the threat posed by quantum computers.}

To prepare for this shift, the National Institute of Standards and Technology (NIST) initiated a standardization effort for quantum-safe digital signatures \cite{noauthor_post-quantum_nodate}. On August 13, 2024, NIST standardized two digital signature algorithms, ML-DSA \cite{fips204} and SLH-DSA \cite{fips205} that are believed to be secure against large-scale quantum computers.

One of the main challenges in adopting these new signatures is their considerable size \cite{adrian_post-quantum_2024}. ML-DSA produces 2,420-byte signatures and SLH-DSA has 7,860-byte signatures, which are orders of magnitude larger than 64-byte Elliptic Curve Digital Signing Algorithm (ECDSA) signatures used today. Embedding such large signatures in tokens would increase transmission costs and latency, and could also make systems susceptible to denial-of-service attacks \cite{adrian_post-quantum_2024, Baseri_2024}. \textcolor{red}{For edge devices such as medical devices, this could make protocols prohibitively computationally expensive and impractical to run.}

Clearly, new authentication and authorization protocols must be designed with these constraints in mind. One such example is KEMTLS \cite{cryptoeprint:2020/534} which replaces the TLS handshake with quantum-safe primitives. Inspired by these developments, and the need for health data security in the operating room (IoT-based surgical systems), we here propose a novel protocol that solves a specific goal: efficient and secure machine-to-machine authorization and authentication in the presence of quantum-capable adversaries. Our protocol uses quantum-safe key encapsulation mechanisms (KEMs) and digital signature algorithms (DSAs) to establish tokens over insecure channels, eliminating the need to include large digital signatures in the tokens themselves. It also supports key rotation. To verify its security properties, we formally analyze the protocol using Tamarin \cite{9768326}, a symbolic verification tool for cryptographic protocols.
\section{Cryptography for Health Data}
Suppose there is a medical device that transmits data to a cloud server. We want only that authorized machine communicating with the cloud. A stronger guarantee is the certainty that the messages sent over the network really \textit{do} originate from the machine, and have not been modified or forged by an attacker.

\IEEEpubidadjcol

\subsection{Integrity}
Suppose the machine has a message $m$ to send to the server which comprises of medical data. This medical data is essential to the treatment of the patient, and thus we must validate that $m$ is received by the server exactly as it was sent by the machine, and not modified maliciously by an attacker.

One cryptographic primitive that could be applied here is a hash function. A hash function $h$ maps strings of arbitrary length $x$ to strings of fixed length $h(x)$. A desirable property of hash functions is \textit{pre-image resistance}. That is, if we have a hash value $h(x)$, it is computationally infeasible to find $x$. We also want \textit{collision resistance}, that is, if we have two unique messages $x_1,x_2$ such that $x_1 \neq x_2$ then $h(x_1)\neq h(x_2)$. \cite{bernstein_post-quantum_2017}

To detect modifications to the message $m$, the machine sends a hash along with the message, making the final message $(m, h(m))$. When the server receives the message, the server computes $h(m)'$ and checks $h(m)'=h(m)$. Since the hash function has collision resistance, if even a single bit was changed, the hash is different and it becomes evident that the message has been modified. However, the obvious attack here is that the message can be modified, the hash recomputed, and in doing so the server is tricked into believing no modification took place.

A digital signature algorithm provides better guarantees. The machine generates a key pair consisting of a private key $k_\text{priv}$ and a public key $k_\text{pub}$. The public key is transmitted publicly so that the attacker and the server can see it. Now, the machine sends the message $(m, \text{Sig}_{k_\text{priv}}(m))$, which is the message $m$ followed by the signature of $m$ under $k_\text{priv}$.

The server uses the public key to trivially verify that $m$ was \textit{indeed} signed by the machine. If $m$ had been modified by an attacker, the signature would no longer be valid, and if the attacker were to modify the signature, then it would no longer be verifiable using the public key. Thus, we have shown that the message was sent by the machine \textit{and} that it has not been modified.

\subsection{Shared Secrets}
\begin{figure}
    \includegraphics[scale=1.0]{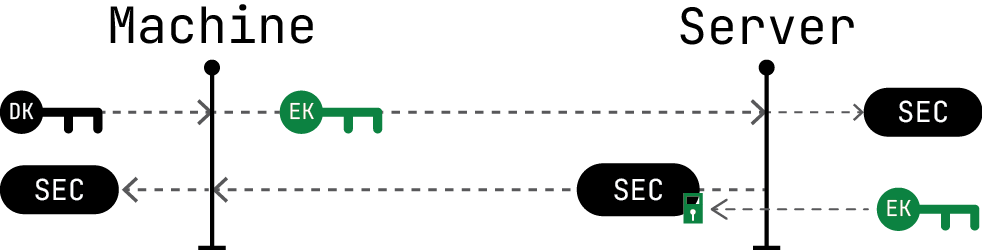}
    \caption{A key encapsulation mechanism (KEM). The client starts with the decapsulation key \texttt{DK} and the encapsulation key \texttt{EK}, sends the \texttt{EK} to the server, which the server uses to encapsulate the shared secret $\texttt{SEC}$ which can be decpasulated by the client. }
    \label{fig:fig_explicatory2}
\end{figure}

Suppose the machine needs to establish a secret string in order to encrypt messages. Considering the initial channel is insecure, a shared secret will need to be established without the attacker learning said secret. For this, a key encapsulation mechanism (KEM) is employed. (Fig.~\ref{fig:fig_explicatory2})

The machine begins by generating a KEM private key, called the decapsulation key $dk$, and then a public key, called the encapsulation key $pk$ which is shared with the server. The server then encapsulates the shared secret $S$ using the encapsulation key to produce the cipher text. The machine receives the cipher text, and uses the decapsulation key to retrieve the secret $S$. All of this occurs without the attacker obtaining knowledge of $S$, and thus a shared secret is successfully established between the two parties.

\begin{figure*}
    \centering
    \includegraphics[scale=0.5]{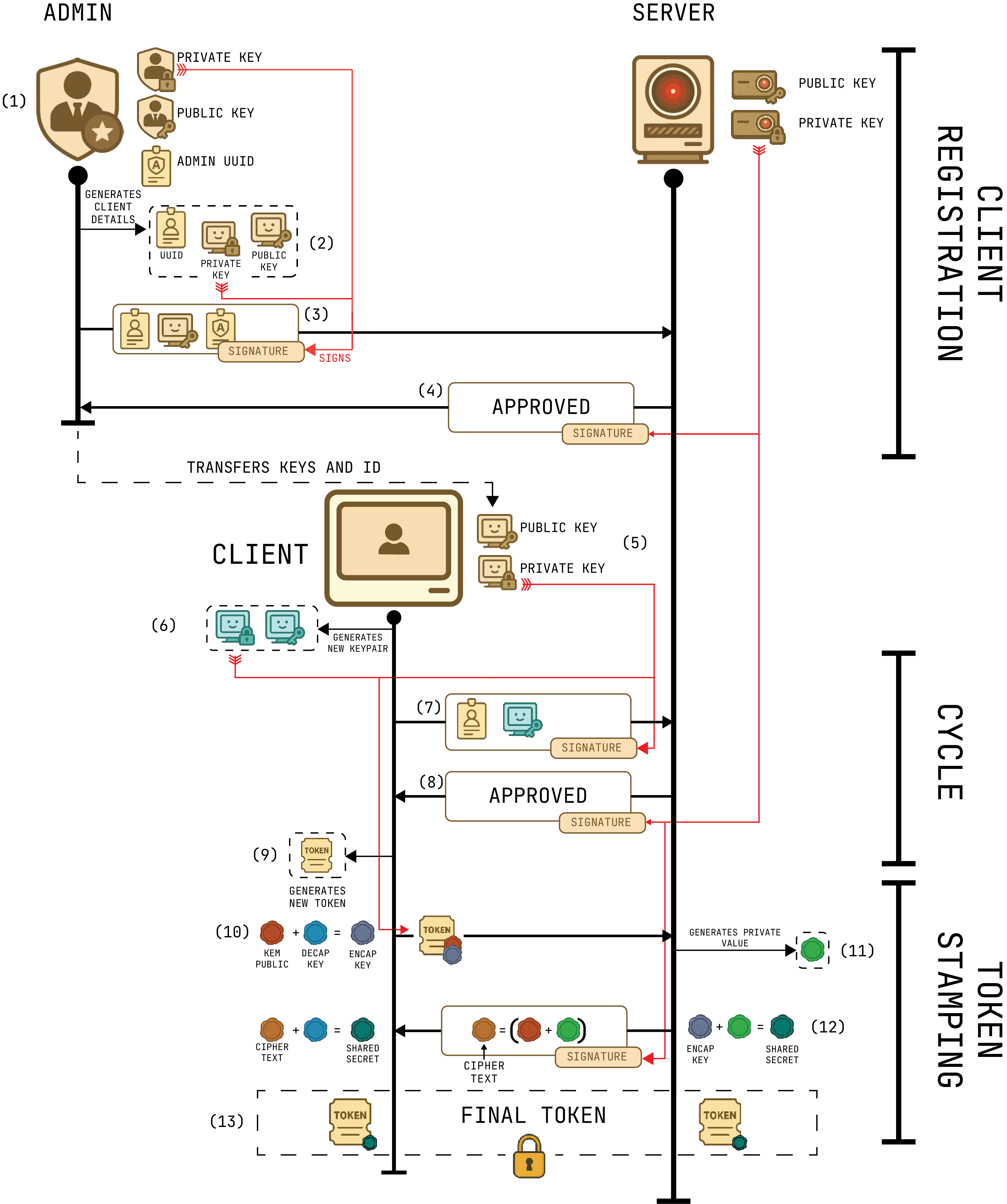}
    \caption{A simplified overview of the protocol. The protocol begins with the administrator (1) generating client credentials (3) and sending them to the server for approval (3). The server approves the credentials (4), and the administrator installs them securely on the client machine (5). To initiate a key cycle, the client generates a new key pair (6) signed with both old and new keys, and sends it to the server (7). The server verifies and approves the key cycle (8). For token stamping, the client generates a preview token (9), attaches a KEM public key and encapsulation key, and transmits this to the server (10). The server uses a private value (11) to generate a ciphertext which is sent to the server (12). Both parties derive the shared secret and compute the final token (13). Cycles and token stamping may take place an arbitrary number of times and in any order.}
    \label{fig:fig_explicatory}
\end{figure*}
\section{Protocol}

\subsection{Overview and Preconditions}
Here we present a brief overview of the protocol (Fig.~\ref{fig:fig_explicatory}) using the same medical device and server from the previous section. \textcolor{red}{We assume that the protocol is carried out over a transport-layer protocol ensuring in-order reliable delivery, such as Transport Control Protocol (TCP).} Our goal is to use quantum-safe digital signatures and KEMs to establish a token to authorize future requests. Before authentication can proceed, the medical device must be registered with the server by an administrator.

The administrator generates an ID for the medical device, along with a private and public key. The administrator forms a message containing the ID and the public key and signs it with both the client key and the administrator key to show that the private key does exist, and that the administrator did authorize this request.

The server will examine and validate this request, then form an approval message and sign it with the server private key. This is received and validated by the administrator. The administrator securely installs the client UUID and private key on the medical device, who will direct all interaction with the server from there on. As the private key serves as the proof-of-identity for the medical device, it is paramount that it is transported securely \cite{ietf-oauth-security-topics-29}.

The medical device may choose to change the key arbitrarily, or when required by the server. This process is called key cycling. The medical device generates a new key pair, and forms a message comprising of the device ID and new public key. This message is signed twice. The first signature is by the new private key, to prove that this key is owned and exists, and the second signature is by the old private key, to authorize this request. This message, along with the signatures, is sent to the server, who trivially validates it, installs the new key, and sends an approval message signed with the server private key.

We will now obtain a token. A preview token is proposed by the medical device which states certain properties that the device would like the token to have. This could be permission scopes or a certain version of the protocol. The token also contains a timestamp and a payload which is a random string of 32 bytes \textcolor{red}{from a cryptographic random number generator.} We also compute a KEM encapsulation \& decapsulation key. The token and encapsulation key are signed and then sent to the server. 

The server validates the signature and the token details, that the timestamp is valid, and that the permission scopes requested are reasonable. The server then uses the encapsulation key to generate a shared secret $\mathfrak{S}$. The token payload is replaced by $\mathfrak{S}$, and then the token is hashed and stored in the server database. The server sends the KEM ciphertext back along with the hash of the final token, signed by the server private key. 

The client uses the KEM ciphertext \& decapsulation key to derive $\mathfrak{S}$, derives the same token, and verifies it by checking the hash. Thus, a secret token is successfully established over an insecure channel.

\subsection{\textcolor{red}{Threat Models}}
\textcolor{red}{As this protocol is designed to be operated over a wireless channel, several new attack vectors are introduced. The protocol is designed to operate over a potentially insecure wireless channel, such as those often found in OR and medical IoT environments, including public hospital wireless networks. In these settings, an adversary may be able to eavesdrop on traffic, inject or modify packets, replay previously observed messages, or attempt to impersonate legitimate parties. These capabilities are consistent with a Dolev-Yao network adversary and are within the scope of this work.}

\textcolor{red}{Certain physical-layer and availability attacks are considered out of scope. For instance, certain physical-layer and availability attacks are considered out of scope, like signal jamming or interference. In this case, the device is prevented from communicating with the server, so messages cannot be exchanged, and no consensus can be established. This, in itself, however, will not compromise any keys exchanged. Power constraints on embedded services, such as those in certain medical devices, motivated the design goal of minimizing signature usage and per-request communication overhead. Within scope, the protocol protects against passive packet sniffing, active message modification, replay attacks, and man-in-the-middle attacks that attempt to alter the protocol state. All security-critical messages are bound to the client or server, respectively, with digital signatures, and the use of a KEM facilitates secure establishment over an insecure channel. As a result, an attacker could not modify permission scopes or timestamps, or substitute key material, without causing signature verification to fail. Replay of previously valid messages is prevented by a monotonic protocol-time counter enforced by both the server and the client. For the purposes of formal analysis and implementation, we assume reliable in-order message delivery (i..e, TCP) and sufficient power for operations during protocol execution. Additionally, the model assumes the authenticity of the server’s public key. The potential mechanisms for bootstrapping server trust are discussed in subsequent sections of the paper.}

\subsection{Protocol Time}
To prevent replay attacks, the protocol uses a special counter-based timestamp known as the \textit{protocol time}. This time stamp system is simple and avoids the need for synchronized clocks or external time sources. \textcolor{red}{This also addresses latency concerns, as heavy latency such as is present in long-distance transmission should not cause timestamps to fail.} The principal goal is to ensure that no two tokens signed under the same key can ever be identical.

Each client maintains its own protocol time with the server, which begins at zero upon successful registration. Every token request, regardless of error or success, increments this value by one. If the client initiates a key cycle, the protocol time is reset to zero. If the client sends an incorrect time, then the server should report back the correct protocol time to resynchronize the counters.

\textcolor{red}{Additionally, the protocol abstracts over retries owing to packet loss, which is why it requires TCP. If a packet is lost, it should be retransmitted identically.}

To prevent timestamp reuse via counter wrapping, the protocol enforces cycling once the maximum value of an unsigned 64- bit integer is reached. This forces the counter back to zero, which is sound since the key has changed and thus old tokens may not be replayed. This design ensures strict monotonicity and protects against replay-based token reuse.

\subsection{Tokens}

\begin{figure}[H]
    \centering
    \begin{bytefield}[bitwidth=0.045em]{90}
        \wordbox{1}{Protocol \\ \tiny 1 byte} &
        \wordbox{1}{Device \\ \tiny 1 byte} &
        \wordbox{1}{UUID \\ \tiny 16 bytes}&
        \wordbox{1}{Perms \\ \tiny 16 bytes} &
        \wordbox{1}{Time \\ \tiny 8 bytes} &
        \wordbox{1}{Payload \\ \tiny 32 bytes}
    \end{bytefield}
    \caption{The layout of the token byte structure in big endian. Each rectangle specifies the contents of the field as well as the amount of bytes the field takes up. Bytes are ordered from top-down, i.e., protocol comes before device type.}
    \label{fig:fig1}
\end{figure}

Tokens have a defined and well-known structure and thus are called \textit{transparent} (Fig~\ref{fig:fig1}). The protocol and device field allow specifying the signing algorithms to use. The UUID field contains the client ID who signed this message.

The 16-byte permission field is for extensibility and is not specified. The timestamp is the protocol time encoded as a 64-bit unsigned integer. The payload is a 32-byte string. In a preview token, this will be random. In a finalized token, this is the shared secret established by the KEM.

\subsubsection{Permission Field} The exact specification of the 16-byte permission field is so that the protocol can be extended, and thus largely it is left out of the specification. The permissions field could also be used to carry basic metadata about the token, or be used to carry permission scopes similar to claims on a JWT\cite{rfc7519}. The first byte of the permission field is reserved for a discriminator, specifying how the permission field will be used.  Modes 0-8 are reserved, and we will specify modes \texttt{0} and \texttt{1} as they cover most cases:
\begin{enumerate}
    \item \textbf{Mode \texttt{0}}: The permission field is turned off and should be disregarded.
    \item \textbf{Mode \texttt{1}}: The permission field represents a $15-$byte code that can be looked up on the server for the scope details.
\end{enumerate}

\subsection{Procedure}

The following sections will describe the protocol (Fig.~\ref{fig:fig_explicatory}) in detail. The network is assumed to be insecure, so an attacker can read and modify any message. However, the protocol could be layered over a secure channel such as TLS. Each section will also briefly discuss the security goals of the various parts of the protocol \cite{provingprotocolscorrect}.

\subsubsection{Special Notation}
For signatures, we write $\text{Sig}_k(m)$ which is the signature of $m$ under $k$. For a private key $k$, the public key is written $\overline{k}$. 

\begin{figure*}[ht]
    \centering
    
    \begin{msc}[instance distance = 5.1cm, head top distance = 0.7cm, left environment distance=1.25cm, right environment distance=1.25cm]{Protocol}
        \declinst{usr}{}{Admin}
        \declinst{m2}{}{Server}
        \declinst{m1}{}{Client}
        \inlinestart{exp1}{ 
  $\quad\quad\text{ }\text{ }$Registration}{usr}{m1}
        \nextlevel
        \nextlevel
        \nextlevel
        \mess[label position=above]{$M_1=\langle \texttt{REGISTER},  I, \overline{k_c}, I_\alpha\rangle$\normalsize}{usr}{m2}
        \nextlevel
        \nextlevel
        \mess[label position=above]{$M_2=\langle \texttt{REGSUCCESS}, h(I)\rangle$}{m2}{usr}
        \nextlevel
        \nextlevel
        \mess[label position=above right, pos=0.6]{$\text{SECURE}(\langle I, k_c\rangle)$}{usr}{m1}
        \nextlevel
        \inlineend{exp1}
        \inlinestart{exp2}{$\quad\quad\text{ }\text{ }$Cycle}{m2}{m1}
        \nextlevel
        \nextlevel
        \nextlevel
        \mess[label position=above]{$M_3 = \langle \texttt{CYCLE}, I, \overline{k'_c}\rangle$}{m1}{m2}
        \nextlevel
        \nextlevel
        \mess[label position=above]{$M_4= \langle \texttt{CYCLEOK}, h(\langle I, \overline{k'_c}\rangle )\rangle $}{m2}{m1}
        \nextlevel

        \inlineend{exp2}

        \inlinestart{exp3}{$\quad\quad\text{ }\text{ }$Token Establishment}{m2}{m1}
        \nextlevel
        \nextlevel
        \nextlevel
        \mess[label position=above]{$M_5= \langle \texttt{STAMP}, \mathfrak{T}, ek) \rangle$}{m1}{m2}
        \nextlevel
        \nextlevel 
        \mess[label position=above]{$M_6 = \langle \texttt{STAMPED}, h(\overline{\mathfrak{T}}, \mathfrak{T}), ct) \rangle$}{m2}{m1}
        \nextlevel

        \inlineend{exp3}

    \end{msc}
    \caption{\textcolor{red}{Protocol Flow. The protocol specification without signatures showing the messages exchanged between the administrator and server and then subsequently between the client and server.  To ensure the text was legible, the signatures have been left out of this diagram, but they are described in detail in the \textit{Procedure} section.}}
    \label{fig:procedure}
\end{figure*}

\subsubsection{Registration}

The goal of registration is that only authorized administrators may register new clients to the server, and that clients who are being registered actually possess the key pair they claim to possess. The exact semantics of administrators is out of scope for this protocol. Registration is initiated by an administrator who possesses the administrator private key $k_\alpha$ and administrator ID $I_\alpha$. The administrator proposes a new client with ID $I_c$ and public key $\overline{k_c}$. The administrator produces the message $M_1= \langle \texttt{REGISTER}, I, \overline{k_c}, I_\alpha\rangle$. The administrator uses the client private key and the administrator private key to produce the final message $(M_1, \text{Sig}_{k_c}(M_1), \text{Sig}_{k_\alpha}(\text{Sig}_{k_c}(M_1)))$, which is sent to the server.

Let the server key pair be $(k_S, \overline{k_S})$.  Upon receiving this message, the server verifies the validity of both signatures and that the public key and identifier are unique. The server produces the message $M_2 = \langle \texttt{REGSUCCESS}, h(I)\rangle$ and then responds with $(M_2, \text{Sig}_{k_S}(M_2))$ to the administrator. The administrator verifies the message was signed by the server, and verifies the identifier hash. This prevents the impersonation of the server, and ensures the registration response may never be replayed since identifiers must be unique.

Through secure means, the administrator installs the client identifier $I$ and private key $k_c$ on the client. All subsequent communications take place between the client and the server.

\subsubsection{Key Cycling}
Key cycling is only necessary when the key expires, as determined by server policy. If a client attempts to get a token with an expired key, the server will direct the client to initiate a cycle. Alternatively, a client may choose to initiate a cycle arbitrarily often.

The goal of key cycling is that only the owner of the private key can update the public key on the server, and that the public key proposed was created by the client and forms part of a legitimate key pair.

To begin the cycling process, the client generates a new key pair $(k'_c, \overline{k'_c})$.  The client then produces the message $M_3 = \langle \texttt{CYCLE}, I, \overline{k'_c}\rangle$ and then sends $(M_3, \text{Sig}_{k'_c}(M_3), \text{Sig}_{k_c}(\text{Sig}_{k'_c}(M_3)))$ to the server. Here we note that the cycling process is practically identical to the registration process, with the difference being that the client self-authorizes, removing the need to include the administrator ID.

The server verifies the signatures, verifies the public key is unique, and computes the verification hash $\mathfrak{v}= h(\langle I, \overline{k'_c}\rangle)$. The server then produces $M_4= \langle \texttt{CYCLEOK}, \mathfrak{v}\rangle$ responds with  $(M_4, \text{Sig}_{k_S}(M_4 ))$. The client verifies the authenticity of the response with the server public key, and verifies the verification hash.

\subsubsection{Token Stamping}

The final operation supported is token establishment, or \textit{stamping}, the central operation of the protocol. Tokens are stamped in such a way to prevent replay attacks. Additionally, since a token verifies against a client, a token is valid if and only if the client itself is valid. Each token stamped corresponds uniquely to a single non-repeating instance of the protocol, fulfilling the property of injective agreement. Tokens are forward-secure due to their reliance on quantum-safe KEMs which are independent from the signing keys. Furthermore, the server stores only hashes of tokens, meaning that an attacker cannot extract valid tokens even in a database breach.

Token establishment process begins with the client generating a preview token. The token has the following structure:
\begin{equation}
    \mathfrak{T}= \langle p_\alpha,p_\beta, I, p_\psi,p_\Delta, p_\zeta\rangle
\end{equation}
where $p_\alpha$ is the protocol, $p_\beta$ the device, $I$ the client UUID, $p_\psi$ the permission field, $p_\Delta$ the timestamp, and $p_\zeta$ the payload field. The payload field may be initialized to a random string of bytes \textcolor{red}{from a cryptographic random number generator.}

The client then generates a KEM encapsulation key $pk$ and decapsulation key $sk$. The client produces $M_5 = \langle \texttt{STAMP}, \mathfrak{T}, pk\rangle$ and then sends $(M_5, \text{Sig}_{k_c}(M_5 ))$ to the server.

The server receives this, verifies the signature, and verifies that the token's timestamp aligns with the server protocol time for that client. Additionally, the server checks and sees that this token is not yet in the database.

The server may also verify the permission scopes, however this is out of scope for the protocol and is left up to the implementer.

The server computes the KEM shared secret $\mathfrak{S}$ and produces the ciphertext $ct$. The server computes the final token $\overline{\mathfrak{T}_S}$ as,
\begin{equation}
    \overline{\mathfrak{T}}_S= \langle p_\alpha,p_\beta, I, p_\psi,p_\Delta, \mathfrak{S}\rangle
\end{equation}
The server then computes the approval hash $\mathfrak{a}_h=h(\langle \overline{\mathfrak{T}_S}, \mathfrak{T}\rangle)$ and produces $M_6= \langle \texttt{STAMPED}, \mathfrak{a}_h, ct\rangle$ sends $(M_6, \text{Sig}_{k_S}(M_6 ))$ to the client.

The client will verify the signature using the server public key. The client then uses the ciphertext $ct$ and the decapsulation key $dk$ to derive the shared secret $\mathfrak{S}$. The client computes the client token:
\begin{equation}
    \overline{\mathfrak{T}}_C= \langle p_\alpha,p_\beta, I, p_\psi,p_\Delta, \mathfrak{S}\rangle
\end{equation}
The client now verifies that $h(\langle \overline{\mathfrak{T}_C}, \mathfrak{T}\rangle)=\mathfrak{a}_h=h(\langle \overline{\mathfrak{T}_S}, \mathfrak{T}\rangle)$, asserting that the token was correctly established and that $\overline{\mathfrak{T}_S}=\overline{\mathfrak{T}_C}$.

\section{\textcolor{red}{Bootstrapping}}

\textcolor{red}{The registration process relies heavily on a secure out-of-band channel to install the client's private key.  This section discusses practical methods for performing the initial bootstrap.}

\begin{enumerate}
    \item \textcolor{red}{\textit{On Manufacture.} The manufacturer performs the registration and installs the initial key on the machine before shipping it to the client.}
    \item \textcolor{red}{\textit{Physical Delivery.} The initial key can be bootstrapped by an administrator delivering the key via a secure medium such as a flash drive.}
    \item \textcolor{red}{\textit{Delivery by Proximity.} A \texttt{base64} encoding of the initial key could be scanned by the device. For example, the administrator might present the QR code at first boot on the device to register the key material. The security of this approach relies on physical proximity and line-of-sight isolation to prevent remote adversaries from intercepting the key.  Similarly, Bluetooth-based bootstrapping could be performed using authenticated pairing modes, which protect against man-in-the-middle attacks in the initial exchange.}
    \item \textcolor{red}{\textit{Trusted Network.} If the current network is trusted and, at the time, is secure from a post-quantum adversary, the initial key could be passed over this network as long as the cycling takes place before the network is no longer post-quantum safe.}
    \item \textcolor{red}{\textit{Quantum Networks.} The relatively new field of quantum networks gives us quantum key distribution protocols such as BB84 \cite{BENNETT20147} which are provably secure under certain conditions and could be used to bootstrap a classical network.}
\end{enumerate}

\section{Analysis}

We formally analyzed the protocol using the Tamarin prover \cite{9768326}, a state-of-the-art tool for symbolic verification of cryptographic protocols. Tamarin supports rigorous security proofs under the Dolev-Yao adversary model \cite{dolevyao} in which the adversary has complete control over the network. The adversary can delete messages, modify them, and has full knowledge of everything sent across the network.

Tamarin includes built-in support for standard cryptographic primitives, such as hashing functions and digital signatures. For instance, hashing is modeled as \texttt{h/1}, which is a unary function with no equations. That is \texttt{h(x1)=h(x2)} if and only if \texttt{x1=x2}. Digital signatures are modeled by the following equations:
\begin{equation}\label{dsaeq}
    \texttt{verify}(\texttt{sign}(m, sk), m, \texttt{pk}(sk))=\texttt{true}
\end{equation}
which states that a valid signature under a private key $sk$ is only verifiable  using its corresponding public key.

Since key encapsulation mechanisms are not natively supported, we incorporate a custom KEM model based on the Tamarin analysis of KEMTLS \cite{cryptoeprint:2020/534}. We introduce three new functions: \texttt{kempk/2}, \texttt{kemencaps/3}, and \texttt{kemdecaps/3} governed by the following equation:
\small\begin{equation}
    \texttt{kemdecaps}(g,\texttt{kemencaps}(g, ss, \texttt{kempk}(g, sk)), sk) = ss
\end{equation}\normalsize
where $g$ represents shared parameters. This captures the core property of KEMs: only the two parties can derive the shared secret. To support protocol-level guarantees such as identity uniqueness and signature verification, we specified formal restrictions, which prune invalid execution traces from the search space. In our case, these are traces where the signature does not verify, or non-unique identifiers are being used.

For the server keys, a minimal public key infrastructure (PKI) was define, allowing us to model the distribution and compromise of the long-term server key pair. Adversarial capabilities were explicitly modeled via rules such as $\texttt{Reveal\_ltk}$, $\texttt{Reveal\_client}$, and $\texttt{Reveal\_token}$, which simulate the leakage of private keys and tokens. Additionally, token expiration was modeled through dedicated rules, enabling the verification of time-bounded access control policies.

Security properties are formalized as lemmas and verified through automated interactive proofs. To demonstrate the protocol's executability in the absence of an attacker, we included several \texttt{exists-trace} lemmas confirming that honest protocol executions are possible.

A current limitation of the model is that each client is restricted to a single key cycle, and tokens can only be verified once per run. While sufficient to validate core security properties, future work should model an unbounded amount of cycles and multiple rounds of token verification to provide full analysis.

The complete Tamarin model and proof scripts are available at:\newline
\href{https://github.com/OrbSurgicalAI/quath}{https://github.com/OrbSurgicalAI/quath}.

\section{Implementation}

\begin{figure}
    \includegraphics[scale=0.60]{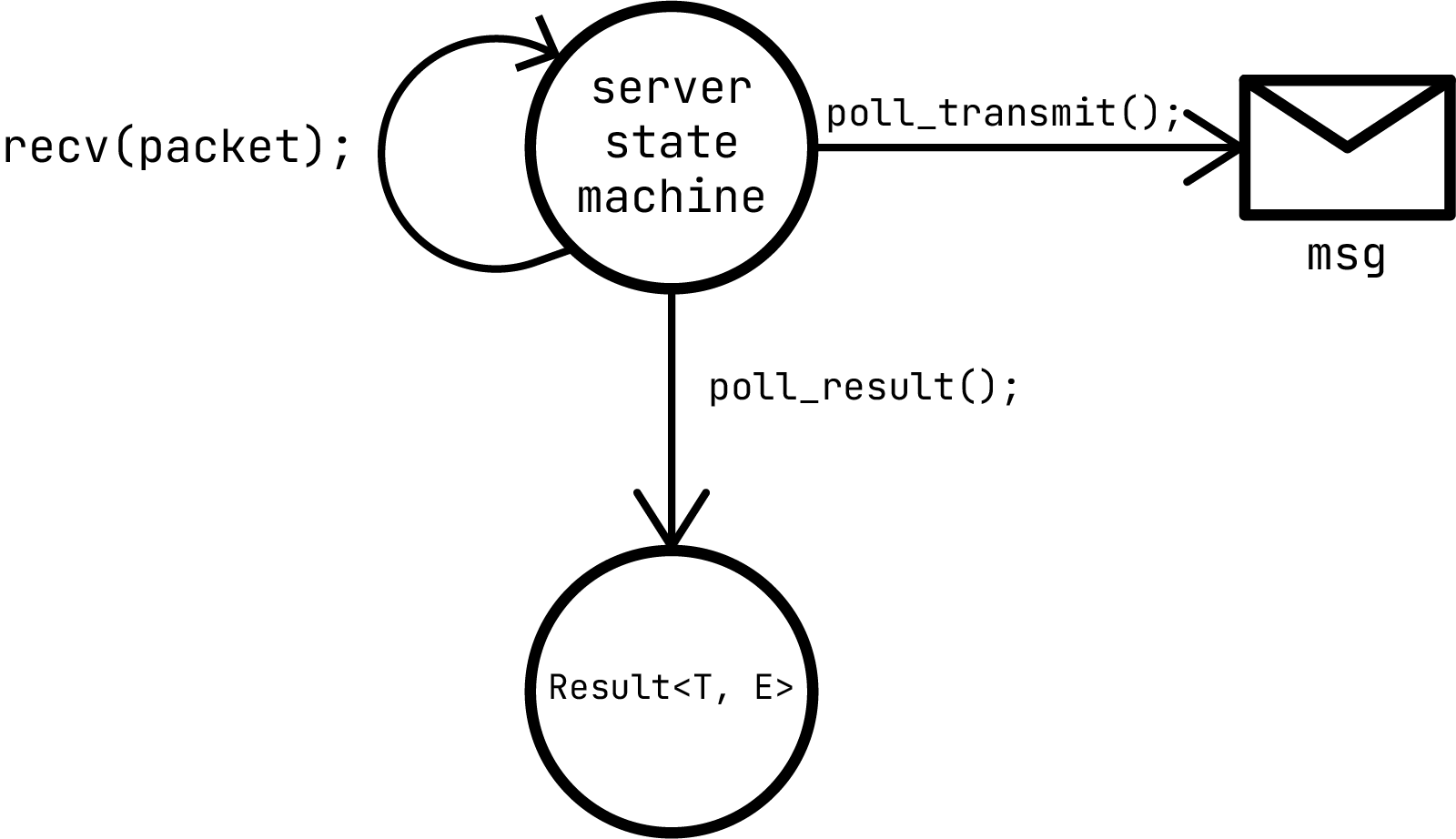}
    \caption{Server state machine in the \texttt{quath} implementation. The \texttt{recv} function processes an input and advances the internal state. The \texttt{poll\_transmit} emits a message if external work is required (i.e., I/O or database interaction), and \texttt{poll\_result} returns either a result or an error.}
    \label{fig:statemachine}
\end{figure}

A core design goal of the protocol was to ensure ease of implementation across a variety of environments, including both synchronous and asynchronous runtimes. To achieve this, we adopted a SANS I/O design pattern \cite{benfield_building_2016} where the protocol logic operates purely on bytes and never performs I/O directly. This decouples the protocol from the execution context, allowing it to be embedded in diverse systems ranging from web servers to embedded devices.

Each component of the protocol (registration, key cycling, and token stamping) is modeled as an explicit state machine (Fig.~\ref{fig:statemachine}). The server state machines are advanced with the \texttt{recv} function, which consumes an input and updates internal state. When external work is required such as I/O or database interaction, the state machine emits a request via \texttt{poll\_transmit}. The driver then invokes \texttt{poll\_result} which can either be \texttt{Pending} if the state machine is awaiting further work, or \texttt{Ready} if the state machine has terminated with a result. Example driver code is included in Fig.~\ref{fig:rustdriving}.

To ensure high performance and type-safety, two qualities which we believe to be critical in safe protocol design, the implementation was written in the Rust programming language. Rust offers strong compile-time guarantees that eliminate entire classes of runtime errors and provides robust memory safety without requiring a garbage collector. These properties make it well-suited for implementing cryptographic protocols where correctness and safety are paramount.

To support flexibility across cryptographic primitives, the implementation leverages Rust's generic trait system to abstract over DSAs, KEMs, and hash functions. This design allows the protocol logic to remain agnostic to specific cryptographic choices, enabling support for a variety of post-quantum algorithms, including ML-DSA \cite{fips204}, SLH-DSA \cite{fips205} and ML-KEM \cite{fips203}.

This modular and generic architecture makes the protocol highly portable and easy to deploy. For instance, an organization such as a bank can readily adopt the system by integrating the state machines into their existing infrastructure. They would simply need to connect the storage and networking handlers, and the protocol will operate seamlessly within the environment.

\EnableColors

\lstset{language=Rust, style=boxed}

\begin{figure}
    \begin{lstlisting}
// Instantiate a State Machine.
let mut machine = StateMachine::new();
// Service queue.
let mut input_queue = VecDeque::new(); 
loop {
     // If we have a new input.
    if let Some(item) = 
        input_queue.pop_back() {
        machine.recv(item);
    }
    if let Some(to_handle) = 
        machine.poll_transmit() {
         // Service the request
         let result = 
            handle_fn(to_handle);
         input_queue
            .push_front(result);
    }
    // The result is ready.
    if let Poll::Ready(result) = 
        machine.poll_result() {
        return result; 
    }
}\end{lstlisting}
    \caption{An example of driving the state machine forward in Rust. In this example the state machine is driven forward by some rust code.}
    \label{fig:rustdriving}
\end{figure}

\DisableColors

\section{Parameters}
\begin{table}[!t]
\caption{\label{tab:table1}%
Recommended protocol parameters based on NIST security levels. Selections follow NIST guidelines.
}
\centering
\begin{tabular}{|c|c|c|c|}
\hline
\textrm{\small Level}&
\textrm{DSA}&
\textrm{KEM}&
\textrm{Hash Function}\\
\hline
1 & ML-DSA-44 & {ML-KEM-512} & SHA3-256\\
\hline 3 & ML-DSA-65 & ML-KEM-768 & SHA3-384\\
\hline 5 & ML-DSA-87 & ML-KEM-1024 & SHA3-512\\\hline 
\end{tabular}
\end{table}

The parameters recommended based on the desired security category are shown in Table~\ref{tab:table1}. The level is the security category based on the weakest link in the parameter set. Digital signature algorithms specified in the table are described in FIPS204 \cite{fips204}, the key encapsulation mechanisms are described in FIPS203 \cite{fips203}, and the hash functions are described in FIPS202 \cite{fips202}.

\color{red}

\section{Performance}

\begin{table}
\caption{\label{tab:table2}Protocol action operation times (ms) at Security Level 1. The values are shown as mean $\pm$ standard deviation over 1000 runs. Client's do not have a verification time as the only cost incurred is transmission, which is left out of this table.}
\label{tab:performance}
\renewcommand{\arraystretch}{1.15}
\setlength{\tabcolsep}{6pt}
\centering
\begin{tabular}{|c|c|c|c|}
\hline
\textbf{Device} & \textbf{Role} & \textbf{Operation} & \textbf{Time (ms)} \\
\hline
\multirow{8}{*}{Raspberry Pi 3}
 & Client & Cycle    & $8.17 \pm 3.09$ \\
 & Client & Register & $8.28 \pm 3.36$ \\
 & Client & Stamp    & $4.47 \pm 2.08$ \\
 & Client & Verify   & N/A \\ \cline{2-4}
 & Server & Cycle    & $4.81 \pm 2.26$ \\
 & Server & Register & $4.82 \pm 2.32$ \\
 & Server & Stamp    & $4.18 \pm 2.22$ \\
 & Server & Verify   & $0.0029 \pm 0.0004$ \\
\hline
\multirow{8}{*}{Raspberry Pi 4}
 & Client & Cycle    & $3.89 \pm 1.43$ \\
 & Client & Register & $3.88 \pm 1.42$ \\
 & Client & Stamp    & $2.19 \pm 1.05$ \\
 & Client & Verify   & N/A \\ \cline{2-4}
 & Server & Cycle    & $2.33 \pm 1.08$ \\
 & Server & Register & $2.29 \pm 1.03$ \\
 & Server & Stamp    & $2.02 \pm 0.98$ \\
 & Server & Verify   & $0.0016 \pm 0.0012$ \\
\hline
\multirow{8}{*}{Intel i7-11800H}
 & Client & Cycle    & $1.03 \pm 0.42$ \\
 & Client & Register & $1.00 \pm 0.38$ \\
 & Client & Stamp    & $0.58 \pm 0.28$ \\
 & Client & Verify   & N/A \\ \cline{2-4}
 & Server & Cycle    & $0.61 \pm 0.29$ \\
 & Server & Register & $0.66 \pm 0.34$ \\
 & Server & Stamp    & $0.54 \pm 0.28$ \\
 & Server & Verify   & $0.0006 \pm 0.0002$ \\
\hline
\end{tabular}
\end{table}

The cost of protocol execution is dominated by the signature algorithm; therefore, a key design goal is to minimize the number of signatures. We would like to highlight that there is a high likelihood that hardware acceleration will be included in future processors, and these times will improve. An additional key design goal of the protocol is that it can operate in resource-constrained environments. Therefore, the protocol was measured on the Raspberry Pi 3 and Raspberry Pi 4. Additionally, for comparison with a user-facing device, a laptop CPU (Intel i7-11800H) was measured, with protocol timings shown in Table~\ref{tab:table2}.

Protocol execution times measure the time required to execute the protocol and do not include transmission overhead, as this is highly dependent on the network. Additionally, to simplify analysis, we have used a standard HashMap as the server's backing store. In a real setup, this could be a database; however, a cache could be used to achieve similar verification times. We observe that even on the Raspberry Pi 3, cycling only takes $\approx 8$ms and stamping takes $\approx 4.5$ms. Furthermore, since cycling occurs only every 90 days and stamping every hour (in a standard configuration), these costs are infrequently incurred, thereby enhancing the efficiency gains.

\section{Overhead Analysis}
In this section we attempt to analyze the communication overhead of our protocol against OAuth 2.0 running in various M2M scenarios. We will focus on the application-layer cost of the protocols, and leave transport costs out of the equation.

\color{black}

\subsection{\textcolor{red}{Our Protocol}}
\textcolor{red}{We begin by describing a general model for determining the average bytes to be expected, regardless of parameters selected in Table~\ref{tab:table1}. As registration occurs once, we restrict the analysis to cycles, token granting, and checking. Let $s_\text{key}$ be the size of the public key for the digital signature algorithm, $s_\text{sig}$ is the size of the signature, $s_\text{hash}$ the size of the hash, $s_\text{ek}$ the size of the encapsulation key, and $s_\text{ct}$ be the size of the KEM ciphertext. Assuming the discriminator (i.e., \texttt{CYCLE}, \texttt{CYCLEOK}, etc.) is a single byte, the key cycle cost is $\mathcal{C}_\text{key}=18+s_\text{key}+3s_\text{sig}+s_\text{hash}$. Then the token renewal cost is $\mathcal{C}_\text{tr}=76 + s_\text{ek}+s_\text{hash}+s_\text{ct}+2s_\text{sig}$.  Finally, the checking cost is just the size of the token, $\mathcal{C}_\text{chk}=74$. Let $\alpha$ be cycles per hour, $\beta$ renewals per hour, and $\gamma$ be token uses per hour. Then we have an expected number of bytes $\mathcal{C}_\text{total}(t)=t(\alpha\mathcal{C}_\text{key}+\beta\mathcal{C}_\text{tr}+\gamma \mathcal{C}_\text{chk})$, where $t$ is a variable representing time in hours. With security level $1$, this would be $\mathcal{C}_\text{total}^\textsf{SEC1}=\alpha 8622+\beta6516+\gamma 74$. Since tokens are typically used far more frequently than they are renewed, the total communication cost is dominated by the term with the coefficient $\gamma$. For constrained devices, communication overhead is often a dominant factor due to limited bandwidth, energy budgets, and stable connection, making compact steady-state tokens particularly advantageous.}

\subsection{\textcolor{red}{Comparison with OAuth 2.0}}

\textcolor{red}{OAuth 2.0 is a broadly specified protocol and has various methods of implementation. Therefore, this section is not meant to give precise communication overheads, but merely recognize the asymptotic gains of this paper and where they are important. The most relevant flow for M2M through OAuth 2.0 is the client credentials grant flow specified in RFC6749 \cite{rfc6749}. One key difference is that OAuth 2.0 explicitly requires a secure channel to obtain an access token, whereas this paper does not. Although OAuth 2.0 is not designed for authentication, protocols that build on top such as OIDC \cite{openidconnect} are designed for user-based authentication. Additionally, OAuth 2.0 does not specify the types of tokens used and this is left to the implementation. We will explore the most common types here. For all methods, we will leave renewal unspecified as it is highly vendor-dependent.}

\subsubsection{\textcolor{red}{JSON Web Signing (JWS) Tokens}}
\textcolor{red}{JWS are the most common form of token and allow signing of JWTs, and incurs communication overhead for the usage of JSON. To be optimistic towards JWS, we assume that they contain the minimum payload for authentication that this paper requires, that is, the 42 bytes. This gives us a theoretical lower bound for the checking term, $\mathcal{C}^\textsf{O-JWS}_\text{chk} \geq 42 + s_\text{sig}$, which means in common setups where the token is used more than it is renewed. Suppose the case where we use ML-DSA at security level one, then $\mathcal{C}_\text{chk}=74$ and $\mathcal{C}^{\textsf{O-JWS}}_\text{chk}= 2462$, or space savings of $97\%$.} 

\subsubsection{\textcolor{red}{JSON Web Encryption (JWE) Tokens}}
\textcolor{red}{JWEs are uncommon and are described in RFC7516 \cite{rfc7516}. At times they themselves contain a JWS, in which case the asymptotic savings are similar to the JWS tokens. However, JWEs also typically use ECDH-ES and RSA, and thus would have to be redefined for the post-quantum case, and therefore they are not suitable for analysis here.}

\color{red}
\subsubsection{\textcolor{red}{Token Introspection}}
Token introspection, as specified in RFC7662 \cite{rfc7662}, enables the use of opaque access tokens whose semantics are resolved through an introspection endpoint. While such tokens themselves can be compact on the wire, this approach requires online validation and as such authorization latency is coupled to the introspection service. This adds additional overhead and latency, and is vendor-dependent and thus difficult to analyze.

\subsubsection{KEMTLS}
While the protocol in this paper and KEMTLS achieve different aims, the KEMTLS implementors do specify the bytes transmitted as part of the protocol \cite{cryptoeprint:2020/534}. If we cache the intermediate certificates and use Kyber and Dilithium (standardized as ML-KEM and ML-DSA, respectively), then the protocol uses $5556$ bytes. In our case, establishment of a token with similar parameters, $\mathcal{C}^\textsf{SEC-1}_\text{tr}=6516$, which is a similar value.

\color{black}

\section{Discussion}

This work presents a practical and formally verified approach for integrating state-of-the-art post-quantum cryptographic primitives into a robust authentication and authorization protocol. The relevance of such systems is increasingly urgent as large-scale quantum computers threaten the long-term security of currently deployed cryptosystems. Critical data, including financial records, medical information, and state secrets, may be vulnerable to attack if post-quantum security is not adopted.

A key feature of our protocol is the compact and extensible token format. Each token occupies only 74 bytes, a stark contrast to the often kilobyte-scale tokens used in comparable systems. This efficiency yields substantial bandwidth savings in high-throughput environments and is particularly advantageous for resource-constrained deployments such as embedded systems.

Beyond efficiency, the protocol offers strong flexibility and extensibility. Tokens can be customized to include permission scopes and application-specific metadata, making the protocol well-suited for secure machine-to-machine interactions in any environment. The protocol need not be limited to machines alone; as enterprise security evolves, this protocol could readily extend to secure user authentication scenarios as well.

A central motivation behind this work is trust, which is a foundational requirement in cryptographic systems \cite{trust10.1145/3511265.3550443}. Trust is especially important when we are dealing with sensitive data such as healthcare data or financial records \cite{NADHAN2024105511}. However, proving the security of protocols remains a profound challenge, which in the past was conjectured to be an insurmountable task \cite{isitpossibletoprovesecure,10.1145/3133956.3134063} due to the inherent complexities in modeling real-world adversarial behaviors. 

Nonetheless, formal verification remains one of the most effective tools for providing provable security guarantees for protocols \cite{isitpossibletoprovesecure, provingprotocolscorrect}.

The need for formal analysis is underscored by the nature of the data guarded by the protocols we use daily. For instance, TLS secures daily web traffic. The triple handshake vulnerability in TLS was uncovered and healed through formal analysis \cite{10.1145/3133956.3134063, 6956559}. Historically, protocols like Otway-Rees \cite{otwayrees}, which appear to be secure on first glance, have been shown to have major flaws with trivial formal analysis. These finings highlight the inadequacy of intuition-driven protocol design and verification, and the necessity of machine-checked verification. 

To that end, we employed the Tamarin Prover \cite{9768326} to formally verify our protocol's key phases: registration, key cycling, and token issuance. Tamarin's symbolic model, employing the Dolev-Yao adversary \cite{dolevyao}, allows reasoning over all possible executions under a network controlled by a powerful adversary. Indeed, Tamarin has been used for proofs of prolific protocols such as TLS1.3 \cite{10.1145/3133956.3134063} and Apple iMessage \cite{cryptoeprint:2024/1395}. 

We prove properties such as injective agreement, session key ownership, and resilience to token expiration attacks under a limited version of the model. Importantly, the model also models scenarios such as database leakage and full-server compromise. These formally proven properties provide robust assurance of the protocol's correctness and resilience, even under a powerful adversary.

Furthermore, it is important to consider how this protocol aligns with compliance in various regions. \color{red}Healthcare regulations such as HIPAA, GDPR, and ISO 27001 impose requirements on confidentiality, integrity, access control, and auditability. Although compliance is achieved at the system level, the protocol was designed to aid in regulatory compliance. The protocol minimizes data; for instance, tokens are fixed-size and do not store personal data. However, implementors must be cautious not to encode personal data within permission scopes or within linked metadata. Permission scopes themselves provide least-privilege access control, and those in the healthcare field should configure permissions so the least-privilege principle is applied. Additionally, explicit token issuance and protocol-time sequencing naturally support audit trails and access logging, facilitating compliance with accountability and monitoring requirements in regulated healthcare environments. This is because service entities have a verifiable identity intrinsically linked to their private key, ensuring authentication via the digital signature algorithm and allowing all access to be logged without revealing the data.

Finally, the protocol is suitable for standardization. It features a compact, well-defined message structure and effectively uses a post-quantum primitive. These properties align well with the requirements typically expected by standards bodies such as IEEE or IETF. In particular, the protocol could be positioned as a secure machine-to-machine authentication in regulated healthcare environments. \color{black}

\subsection{Limitations}
One limitation of the protocol is that the client must have prior knowledge of the server's public key. \color{red}This requirement can be mitigated by executing the protocol over an authenticated secure channel, such as one established through KEMTLS, thus eliminating the necessity to pre-distribute the server's public key. For simpler alternatives, a trust-on-first-use (TOFU) model could be applied with key-pinning for the server key. In cases where the initial bootstrap is done via proximity or physical interaction, the server key could also be installed simultaneously.  Additionally, future work could investigate extensions toward running the protocol in a mutual mode, where each client could also act as a server. \color{black} Additionally, features such as token revocation could be explicitly modeled in future versions of the protocol.

\color{red} A second limitation concerns the scope of the formal security analysis rather than the protocol's design. Although the protocol is intended for long-lived devices and supports repeated key cycling and token establishment, the current Tamarin model verifies these properties for a single protocol cycle and a single round of token verification. This restriction was introduced to control state-space explosion during symbolic analysis, ensuring that the analysis remained tractable. Accordingly, the present work does not claim full multi-round or unbounded key-cycle security proofs.

Extending the formal model to cover an arbitrary number of key cycles and repeat token verification is an important direction for future work. Symbolic verification of protocols with cyclic states remains an active research area, and recent advances in Tamarin support enhanced inductive reasoning over repeated protocol executions \cite{10.1145/3719027.3765131}. Future work will leverage these techniques to provide multi-round, inductive proofs that further strengthen confidence in the protocol's long-term security assurances.

\color{black}

\subsection*{Additional Information}

\textbf{Contact Information:} Homer A. Riva-Cambrin, homer.rivacambrin@ucalgary.ca; Sanju Lama, slama@ucalgary.ca; Garnette R. Sutherland, garnette@ucalgary.ca; Rahul Singh, rahul.singh3@ucalgary.ca;

\textbf{Competing Interests:} The security protocol as presented in the work, and its widespread application, stems from the need for health data protection relative to our innovations in the Operating Room technologies. All authors have affiliation to a University of Calgary (Project neuroArm Medical Robotics research facility) spin-off called OrbSurgical Ltd.

\textbf{Funding:} Canadian Institutes for Health Research Commercialization Grant Application \#390405 .

\textbf{Data Sharing:} All code and data can be located at the following link: \newline\href{https://github.com/OrbSurgicalAI/quath}{https://github.com/OrbSurgicalAI/quath}

\bibliographystyle{IEEEtran}
\bibliography{pnas-sample}

\vfill

\end{document}